\documentstyle [12pt] {article}
\begin{document}
\title{From the paradoxes of the standard wave-packet analysis to
the definition of tunneling times for particles}

\author{N.L.Chuprikov \\ Tomsk State Pedagogical University \\ 634041
Tomsk, Russia} \date{\today}

\maketitle

\begin{abstract}

We develop a new variant of the wave-packet analysis and solve the
tunneling time problem for one particle. Our approach suggests an
individual asymptotic description of the quantum subensembles of
transmitted and reflected particles both at the final and initial stage
of tunneling.  We find the initial states of both subensembles, which
are non-orthogonal. The latter reflects ultimately the fact that at the
initial stage of tunneling it is impossible to predict whether a
particle will be transmitted through or reflected off the barrier. At
the same time, in this case, one can say about the to-be-transmitted
and to-be-reflected subensembles of particles. We show that before the
interaction of the incident packet with the barrier both the number of
particles and the expectation value of the particle's momentum are
constant for each subensemble. Besides, these asymptotic quantities
coincide with those of the corresponding scattered subensemble. On the
basis of this formalism we define individual delay times for both
scattering channels. Taking into account the spreading of wave packets,
we define the low bound of the scattering time to describe the whole
quantum ensemble of particles.  All three characteristic times are
derived in terms of the expectation values of the position and momentum
operators. The condition which must be fulfilled for completed
scattering events is derived.  We propose also the scheme of a {\it
gedanken} experiment that allows one to verify our approach.

\end{abstract}

\newpage

\section*{Introduction}

The tunneling time problem is perhaps one of the most long-lived and
involved problems of quantum mechanics. Although there exists a variety
of proposals (see, for example, reviews \cite{Ha2, La1, Mu0}) to define
characteristic times for a tunneling particle, this question have been
continued to be controversial. The first attempts (see, for example,
\cite{Ha2, Ha1, Le1, Ter, Col}) to solve this problem, on the basis of
the one-dimensional Schr\"odinger equation (OSE), were reduced to
the naive analysis of the relative motion of the transmitted (or
reflected) and incident wave packets. Such an approach is usually
referred to as the standard wave-packet analysis (SWPA). The main
result of the SWPA is the development of concepts of the phase and
delay times to describe particles with a definite momentum.  However,
even in this particular case these times proved to be ill-defined.
Their rigorous justification leads to the difficulties which remain to
be overcome (see \cite{Ha2, La1}).  But the most serious difficulties
in the SWPA arose in attempt to define tunneling times for wave packets
whose width is comparable with the width of a potential barrier. Many
authors (see, for instance, \cite {Ha2, La1}) pointed out that some
properties of tunneling such wave packets (the dependence of the phase
and delay times on the initial distance between the incident wave
packet and the potential barrier; the acceleration of the transmitted
wave packet by an opaque potential barrier) seem to be paradoxical. Due
to this fact the SWPA has been considered (see \cite{La1, Mu0}) as the
most unpromising approach to the solution of the tunneling time
problem. As a result, all the significant papers published on this
problem for the last ten years have been fulfilled out of the framework
of the SWPA (see reviews \cite{La1, Mu0} and also \cite{Gru, Mu1}).
However, as is seen from \cite{La1, Mu0}, the alternative approaches
themselves meet with the serious difficulties.  The basic assumptions
(to prescribe the Feynman, Bohmian or Wigner trajectories to a
particle; to incorporate in quantum mechanics the conditional
probabilities which may be negative by value; to introduce the special
"quantum" clocks for timing the particle's motion; to admit the
averaging procedures (for example, averaging over time intervals) to
differ from the averaging over a quantum ensemble; to admit making use,
in addition to parameter $t$, of operators with the dimension of time)
underlying these approaches are either purely speculative or
questionable in many respects. In any case, such assumptions lead the
alternative approaches out of the framework of the conventional quantum
mechanics and, as a consequence, they themselves need an independent
justification. A detailed analysis of this problem can be found in
\cite{La1, Mu0, Ste} (see also \cite{Mu4}).

We have also to add that none of them explains the mentioned above
paradoxical behavior of wave packets. The point is that such an
explanation have not merely been a purpose of the alternative
approaches. As was said by Landauer and Martin \cite{La1}, this is
"...a responsibility of wave-packet analysis proponents...". However,
in our opinion, such a viewpoint is at least strange because the
description of the particle tunneling in terms of wave packets is
inherent to quantum theory. So that an exhuastive interpretation of the
wave-packet dynamics in tunneling is a task not only for the proponents
of the SWPA. All approaches whose aim is to solve the tunneling time
problem must explain the behavior of wave packets in tunneling.

Taking into account the above, we see a necessity to return to the
SWPA, and clarify the reasons having led to the failure of the earlier
approaches in solving the given problem.  We begin our analysis with
the remark that there is really the case in the SWPA when no problems
arise in timing the particle's motion. We have in mind the motion of an
everywhere free particle. Namely the timing procedure for a free
particle was used in defining the phase times (see \cite{Ha2, Ha1, Le1,
Ter}). The possibility to use such a procedure in the tunneling problem
is based on the fact that a tunneling particle moves freely before and
after its interaction with the potential barrier.

To avoid some mistakes in the more complex, scattering case, one has to
dwell shortly on the free-particle timing procedure. A simple analysis
of classical dynamics of particles shows that it is important to
distinguish two kinds of characteristic times: 1) the time of arrival
of a free particle at some given point, and 2) the duration of passing
the particle through some given spatial interval. Defining the first
quantity is based in quantum mechanics on the fact that the expectation
values of the velocity and position of a free particle coincide with
the corresponding characteristics of the "center of mass" (CM) of a
wave packet to describe the particle state. As is known, in this case
the motion of the CM is described by Newton equations for a free
classical particle.  Thus, in order to define the (mean) time of
arrival of a quantum particle at a given point, one needs simply to
follow the CM of the corresponding wave packet. This concept will be
used here in defining delay times.

A more distinct difference between the classical and quantum dynamics
of free particles arises in determining the characteristic times of the
second kind.  So, the time a particle spends within some spatial region
as well as the time the particle spends to traverse this region are the
same classically. However, this is not the case in quantum mechanics.
For example, a quantum particle spends no time in the null spatial
interval. But there is a non-zero temporal interval when the
probability for a particle to traverse this region is large enough. To
define the duration of the physical process, one has to follow at least
the fronts of a wave packet, rather than its CM. For this purpose, in
addition to the expectation value of the position operator at some
instant $t$, one needs to know its mean-square deviation. This concept
will be used below in defining the low bound of the scattering time.

The free-motion case reveals the following two aspects of timing a
quantum dynamics of particles. Firstly, timing the motion of a quantum
particle can be thought only as timing the evolution of different
(initial or central) moments of the position (or another Hermitian)
operator.  This is related ultimately to the fact that quantum
mechanics can predict only expectation values of physical quantities.
Secondly, there is no necessity to construct special "quantum clocks"
for timing the motion of a free quantum particle.  Characteristic times
for such a particle (in other words, for the CM or fronts of a free
wave packet) can be measured with the help of the same clocks that used
in keeping track of a classical free particle.

It is obvious that the above peculiarities of timing a free particle
must be true also for a scattering particle when it moves beyond the
scattering region (this suggests that the scattering event is a
completed scattering). However, in this case a new aspect of timing
arises: while a particle interacts with the barrier, to time its motion
is impossible.  There are two main reasons for this.  Firstly, while a
wave packet interacts with the barrier, the position and velocity of
its CM say nothing about the expectation values of the position and
velocity of a particle: because of the interference, in this case they
can be unambiguously related neither to an incident nor transmitted,
nor reflected particle. Secondly, any measurement in the barrier
region contradicts the conventional setting of the tunneling problem.

As is seen, the free-particle timing procedure itself is simple enough.
However, the application of this procedure to a scattering particle, in
the case when the wave packet width is comparable with the barrier
width, leads to the ill-defined phase times (see, for example,
\cite{Ha2, La1, Ha1, Nas, Ja1}. And the formalism based on this
procedure does not provide a proper explanation to the behavior of wave
packets in tunneling. As is known, a numerical modelling of the
tunneling process for wide (in $x$-space) wave packets shows that,
though the interaction of a particle with the barrier is elastic, the
CMs of the incident, transmitted and reflected wave packets move with
the different velocities. As a rule, the transmitted packet moves
faster than the incident one (for example, it takes place for an opaque
rectangular barrier). In this case the CM of the transmitted packet
appears behind of the barrier well before the incident packet arrives
at the barrier region, i.e., before forming the reflected packet. The
corresponding phase times were found to depend on the initial distance
between the barrier and wave packet: the farther from the barrier is at
the initial time the CM of the incident packet, the earlier the CM of
the transmitted packet appears behind of the barrier.

Thus, according to the SWPA, in the general case the transmitted
particles seem to be accelerated (on the average) in tunneling, and
the corresponding delay time (positive or negative) depends strongly
on the initial distance between the particle and barrier.  Note that
the observed behavior of wave packets is entirely in accordance with
the foundations of quantum mechanics, since it follows from the OSE to
describe the tunneling process. At the same time, its interpretation
mentioned above appears to be mistaken. It is intuitively evident
that a static potential barrier must not accelerate (on the average) a
particle, wherever it moves after the scattering event. Besides, the
delay time must not depend on the initial distance between the
particle and barrier. So that the behavior of wave packets in tunneling
requires a reasonable explanation. The lack of such an explanation as
well as the lack of well-defined characteristic times have a common
root in the SWPA.

At first sight, all the above means that the timing procedure for
an everywhere free particle cannot be, in principle, used for a
tunneling particle.  However, in our opinion, this is not the case.
One has only to bear in mind that the incident, transmitted and
reflected wave packets describe quantum ensembles with the different
number of particles.  This fact have not been taken into account in the
SWPA. The phase time for transmitted particles was defined in this
approach as the difference between the time of arrival of the CM of
the transmitted packet at a point far behind of the barrier, and that
of the CM of the incident packet at a point far before the barrier.
Several authors have been cast doubt on the correctness of such a step.
For example, considering the behavior of the incident and transmitted
wave packets in the numerical modeling of tunneling, the authors of
\cite{La2, La3} emphasize that "...  arriving peaks do not turn into
transmitted peaks ...". In \cite{Ha2, Ha1, Ter} it is pointed to "...
the desirability of constructing an "effective" incident packet which
would play the role of the counterpart to the transmitted packet..."
Moreover, as was pointed out in \cite{Ja1}, classical mechanics"...
would suggest that the energy distribution of the transmitted and
reflected particles is the same before and after the collision with the
barrier..."

To take into account these remarks, one needs to decompose the incident
wave packet into (to-be-transmitted and to-be-reflected) parts.
However, there have been an opinion that this step would contradict the
basic principles of quantum theory. For example, Jaworski and Wardlaw
\cite{Ja1} pointed out "...  that quantum mechanically there is no
sense in speaking about transmitted and reflected particles before they
are detected as such..." As a result, the above idea remains unrealized.
All the known attempts to describe individually transmitted and
reflected particles have been made beyond the idea of the individual
description of these subensembles at the initial stage of scattering.
For example, Steinberg \cite{Ste} offered to introduce the so-called
conditional probabilities (however, being complex, these quantities
lead to the serious problem of their interpretation). There is also a
radical viewpoint (see, for example, \cite{Nus}) that an individual
description of transmitted and reflected particles is inadmissible in
quantum mechanics.

However, as will be shown below, in the frame of the statistical
interpretation (SI) of quantum mechanics (see, for example,
\cite{Bal}), the desirable decomposition does not contradict quantum
theory. Besides, there is another aspect of the tunneling problem that
also requires such a decomposition. We have in mind making use of the
term "two-channel scattering" in solving the tunneling time problem.
So, according to the definition given in \cite{Gol,Tei}, this elastic
one-particle one-dimensional scattering must be treated as an
one-channel scattering. At the same time, according to Hauge and St\o
vneng \cite{Ha2}, the term "scattering channel" in this problem should
be associated with "...  distinct final states (transmission and
reflection ...)".  By this definition, tunneling is a two-channel
scattering.

We agree entirely with this viewpoint. The mere fact that investigators
attempt to define characteristic times individually for a transmitted
and reflected particle means that they distinguish in this task two
scattering channels.  However, one has to recognize that such a
viewpoint needs a proper theoretical basis.  Namely, one should to
develop a formalism to describe individually the quantum subensembles of
transmitted and reflected particles both at the initial and final
stages of tunneling.  Strictly speaking, it is illegitimately to say,
without such a formalism, about two scattering channels in this case.
In our opinion, just ignoring this fact leads to the above paradoxes.

Note that there is a variety of scattering problems related to the
electron transport (see, for example, \cite{Cap}) in artificial quantum
structures (e.g., in quantum wires) where a similar situation takes
place. Right quantum wires which are nonuniform in some limited spatial
interval give the most simple generalization of the considered
one-dimensional model.  More complicated scattering problems arise for
branching quantum wires where the number of scattering channels is
larger than two. It is obvious that the above difficulties of the
one-dimensional case must appear in these models too. To overcome them,
one needs to treat the one-dimensional scattering as a two-channel one,
and develop a relevant theory.

So, in our opinion, the theory of tunneling must include an individual
asymptotic description of the subensembles of transmitted and reflected
particles. The purpose of our paper is to elaborate a desirable
formalism in the framework of the SI (in our opinion, solving the
tunneling time problem is sensitive to the choice of some
interpretation of quantum mechanics). Note that our ideas to solve this
problem have been offered in papers \cite{Ch4, Ch3}. However some
aspects of that solution must be corrected and developed further. In
particular, we have to study here the role of spreading wave packets,
which have been leaved out of account in \cite{Ch4, Ch3}. This effect
is obvious to increase the scattering time. Moreover, when the velocity
of the CM is smaller than that of the wave-packet's fronts, tunneling
represents an incompleted scattering.

The paper is organized as follows. In Section 1 we set the tunneling
problem. In Section 2 we display explicitly the paradoxes arising in
the standard approaches. For this purpose we analyze in detail the
behavior of wave packets in the context of the SWPA. In Section 3 we
show that the wave function describing the quantum ensemble of
tunneling particles contains all information needed for the individual
description of both scattering channels at the initial stage of
scattering. We find here the corresponding counterparts for the
transmitted and reflected packets, which describe individually both
scattering channels at early times.  In this section we answer also the
question, how, in principle, to verify experimentally the formalism
presented. In Section 4 we define delay times for transmitted and
reflected particles. To estimate the duration of the scattering event,
in Section 5 we determine the low bound of the scattering time.
Besides, we derive here the condition which must be fulfilled for a
completed scattering. In Section 6, the properties of the
characteristic times are illustrated in the case of the Gaussian wave
packet tunneling through a rectangular barrier. Some useful
expressions for the asymptotic expectation values of the position and
wave-number operators as well as for their mean-square deviations are
presented in Appendix.

\newcommand{\ko}{\kappa_0^2}
\newcommand{\kj}{\kappa_j^2}
\newcommand{\kd}{\kappa_j d_j}
\newcommand{\kki}{\kappa_0\kappa_j}

\newcommand{\Ra}{R_{j+1}}
\newcommand{\Rb}{R_{(1,j)}}
\newcommand{\Rc}{R_{(1,j+1)}}

\newcommand{\Ta}{T_{j+1}}
\newcommand{\Tb}{T_{(1,j)}}
\newcommand{\Tc}{T_{(1,j+1)}}

\newcommand{\Wa}{w_{j+1}}
\newcommand{\Wb}{w_{(1,j)}}
\newcommand{\Wc}{w_{(1,j+1)}}

\newcommand{\UU}{u^{(+)}_{(1,j)}}
\newcommand{\VV}{u^{(-)}_{(1,j)}}

\newcommand{\ta}{t_{j+1}}
\newcommand{\tb}{t_{(1,j)}}
\newcommand{\tc}{t_{(1,j+1)}}

\newcommand{\tee}{\vartheta_{(1,j)}}

\newcommand{\tta}{\tau_{j+1}}
\newcommand{\ttb}{\tau_{(1,j)}}
\newcommand{\ttc}{\tau_{(1,j+1)}}

\newcommand{\FF}{\chi_{(1,j)}}
\newcommand {\aro}{(k;k_0,l_0)}

\section{Setting the problem for a completed scattering}

Suppose that a particle moves from the left toward the time-independent
potential barrier $V(x)$ which confined to the finite spatial interval
$[a,b]$ $(a>0)$; $d=b-a$ is the barrier width. Let the state of a
particle be described at $t=0$ by the normalized wave function
$\Psi_0(x)$ belonging to the set $S_{\infty}$ which consists from
infinitely differentiable functions vanishing exponentially in the
limit $|x|\to \infty$. The Fourier-transforms of such functions are
known to belong to the set $S_{\infty}$ as well.  This property
guarantees that the position and momentum operators both are
well-defined. Let $<\Psi_0|\hat{x}|\Psi_0>=0,$ $<\Psi_0|\hat{p}|\Psi_0>
=\hbar k_0 > 0,$ $l_0^2=<\Psi_0|\hat{x}^2|\Psi_0>;$ here $l_0$ is the
wave-packet's half-width at $t=0$ ($l_0<<a$); $\hat{x}$ and $\hat{p}$
are the operators of the particle's position and momentum,
respectively.

Another important requirement should restrict the rate of spreading
the incident wave packet. We must be sure that at early times all
particles of the corresponding quantum ensemble move toward the barrier
(only a negligible part of particles is assumed to move at $t=0$ away
from the barrier). This means that the packet's spreading must be
sufficiently ineffective (see condition (\ref{304}) in Section 5).

For our purposes it is useful also to introduce the following
notations.  Let $\hat{H}$ be the Hamiltonian of the problem:
$\hat{H}=\hat{H}_0+V(x)$, where $\hat{H}_0$ is the energy operator for
a free particle.  Besides, let $\hat{H}_0^{ref}$ be the auxiliary
Hamiltonian describing the reflection of a free particle off the
absolutely opaque potential wall located at the middle point of the
interval $[a,b]$, i.e., at $x_{midp}=(a+b)/2$. Let us introduce also
the projection operators $\hat{\Theta}_+$ and $\hat{\Theta}_-$
coinciding with the corresponding $\theta$-functions:  $\hat{\Theta}_+
=\theta(x-b)$, $\hat{\Theta}_- =\theta(a-x)$.

In the general form, the solution of the corresponding temporal OSE can
be written formally as $e^{-i\hat{H}t/\hbar}\Psi_0(x).$ To solve
explicitly this equation, we will use the transfer matrix method (TMM)
\cite{Ch1} that allows one to calculate the tunneling parameters for
any system of potential barriers. Then, the (matched) solution
describing the incident and reflected waves ($x<a$) with the given
wave-number $k$ can be written as

\begin{equation} \label{1}
\Psi_{left}=\left[\exp(ikx)
+\phi_{ref}(k)\exp(-ikx)\right]\exp[-iE(k)t/\hbar],
\end{equation}

\noindent where

\[\phi_{ref}(k)=\sqrt{R(k)}\exp\left[i(2ka+J(k)-F(k)-
\frac{\pi}{2})\right];\]

\noindent the solution

\begin{equation} \label{2}
\Psi_{right}=\phi_{tr}(k)\exp[i(kx-E(k)t/\hbar)],
\end{equation}

\noindent represents the transmitted wave ($x>b$);

\[\phi_{tr}(k)=\sqrt{T(k)}\exp[i(J(k)-kd)];\]
\noindent Here $E(k)=\hbar^2 k^2/2m $; $T(k)$  (the real transmission
coefficient) and $J(k)$ (phase) are even and odd functions of $k$,
respectively; $F(-k)=\pi-F(k)$.

Thus, for the temporal OSE, the solution to satisfy the above initial
condition is given by

\begin{equation} \label{3}
\Psi_{left}(x,t)=\frac{1}{\sqrt{2\pi}}\int_{-\infty}^{\infty}
\left[f_{inc}(k,t) +f_{ref}(k,t)\right]\exp(ikx)dk
\end{equation}

\noindent (for the region $x<a$)
\noindent and

\begin{equation} \label{4}
\Psi_{right}(x,t)=\frac{1}{\sqrt{2\pi}}\int_{-\infty}^{\infty}
f_{tr}(k,t)\exp(ikx)dk
\end{equation}

\noindent (for $x>b$), where

\[f_{inc}(k,t)=c\cdot A\aro\exp[-iE(k)t/\hbar],\]
\[f_{ref}(-k,t)=\phi_{ref}(k) f_{inc}(k,t),\]
\[f_{tr}(k,t)=\phi_{tr}(k) f_{inc}(k,t);\]
$c$ is a normalization constant; the weight function $A\aro$ is the
Fourier-transform of $\Psi_0(x)$. By the initial condition both
$\Psi_0(x)$ and $A\aro$ belong to $S_{\infty}$.  In particular, for
$\Psi_0(x)$ describing the Gaussian wave packet,
$A\aro=\exp\left(-l_0^2 (k-k_0)^2\right)$.

Here functions (\ref{3}) and (\ref{4}) present wave packets moving in
the out-of-barrier regions. Expression (\ref{4}) describes the
transmitted wave packet. Solution (\ref{3}) consists of the incident
and reflected packets (in fact, the latter appears only on arriving the
incident packet at the barrier region).  To study these packets, it is
convenient to pass into the $k$-representation. It is obvious that in
the case of a completed scattering, at late times, the functions
$f_{tr}(k,t)$ and $f_{ref}(k,t)$ approximate Fourier-transforms of the
functions $\hat{\Theta}_+ e^{-i\hat{H}t/\hbar}\Psi_0(x)$ and
$\hat{\Theta}_- e^{-i\hat{H}t/\hbar}\Psi_0(x),$ respectively. The
function $f_{inc}(k,t)$ approximates the Fourier-transform of
$e^{-i\hat{H}t/\hbar}\Psi_0(x)$ at early times. A detailed description
of these packets in the $k$ representation is presented in Appendix.

As follows from Appendix (see (\ref{207}), (\ref{209}) and
(\ref{210})), for a completed scattering the following normalization
conditions must be valid (see also \cite{Ja1}).
\noindent For early times

\begin{equation} \label{70}
\int_{-\infty}^a|\Psi_{left}(x,t)|^2dx \approx
\int_{-\infty}^\infty |f_{inc}(k,t)|^2dk=1;
\end{equation}

\noindent for sufficiently large times
\begin{equation} \label{71}
\int_{-\infty}^a|\Psi_{left}(x,t)|^2dx \approx
\int_{-\infty}^\infty |f_{ref}(k,t)|^2dk=\bar{R};
\end{equation}

\begin{equation} \label{72}
\int_{b}^\infty|\Psi_{right}(x,t)|^2dx\approx
\int_{-\infty}^\infty |f_{tr}(k,t)|^2dk=\bar{T}.
\end{equation}
Here $\bar{T}$ and $\bar{R}$ are the mean values of the transmission
and reflection coefficients, respectively: $\bar{T}=<T(k)>_{inc}$;
$\bar{R}=<R(k)>_{inc}$; angle brackets denote averaging over a relevant
wave packet (see (\ref{205})). We will assume further that conditions
(\ref{70})-(\ref{72}) are satisfied.

As was pointed out above, timing a particle can be made only beyond the
scattering region. All needed information about the influence of the
potential barrier on a particle is contained, via the tunneling
parameters, in expressions (\ref{3}) and (\ref{4}). We will consider
that these parameters have already been known.

\section {Paradoxes of the standard wave-packet analysis}

In order to display explicitly the basic shortcoming of the SWPA, let
us rederive the tunneling times in the context of that approach. The
only difference is that we use here the TMM \cite{Ch1}.

As is shown in Appendix, the expectation values of the $\hat{x}$-operator
for all three packets are given by expressions (see (\ref{216}) -
(\ref{218}))

\begin{equation} \label{5}
<\hat{x}>_{inc}(t)=m^{-1}\hbar k_0t
\end{equation}
for the sufficiently early times, and

\begin{equation} \label{6}
<\hat{x}>_{tr}(t)=m^{-1}\hbar <k>_{tr}t+d-<J^\prime>_{tr},
\end{equation}

\begin{equation} \label{7}
<\hat{x}>_{ref}(t)=-m^{-1}\hbar <-k>_{ref}t+2a+
<J^\prime -F^\prime>_{ref}
\end{equation}

\noindent for the sufficiently large times. Note (see Appendix) that
the prime denotes here the derivative with respect to $k$.

Let $Z_1$ be a point to lie at some distance $L_1$ ($L_1\gg l_0$ and
$a-L_1\gg l_0$) from the left boundary of the barrier, and $Z_2$ be a
point to lie at some distance $L_2$ ($L_2\gg l_0$) from its right
boundary. Thus, conditions (\ref{70}) - (\ref{72}) are fulfilled in
this case. Note that the fulfilment of the condition $a-L_1\gg l_0$
must guarantee that (almost) all particles of the quantum ensemble are
localized, at $t=0$, to the left of the detector placed at the point
$Z_1$. In any case, a part of the quantum ensemble to violate this
condition must be much smaller than $\bar{T}$.

Following the SWPA \cite{Ha2, Ha1, Le1, Ter}, let us define the
difference between the time of arrival of the CM of the incident packet
at the point $Z_1$, and that of the transmitted packet at the point
$Z_2$ (this time will be called below as the "transmission time").
Analogously, let the "reflection time" be the difference between the
time of arrival of the CM of the incident packet at the point $Z_1$ and
that of the reflected wave packet at the same point.

\newcommand{\ppp}{\mbox{\hspace{10mm}}}
\newcommand{\ooo}{\mbox{\hspace{3mm}}}

Let $t_1$ and $t_2$ be such instants of time
that

\begin{equation} \label{8}
<\hat{x}>_{inc}(t_1)=a-L_1; \ppp
<\hat{x}>_{tr}(t_2)=b+L_2.
\end{equation}

\noindent Considering (\ref{5}) and (\ref{6}), one can write then the
"transmission time" $\Delta t_{tr}$ ($\Delta t_{tr} =t_2
-t_1$) for the given interval in the form

\begin{equation} \label{9}
\Delta t_{tr}=\frac{m}{\hbar}\left[\frac{<J^\prime>_{tr} +L_2}
{<k>_{tr}} +\frac{L_1}{k_0}+ a\left(\frac{1}{<k>_{tr}}
-\frac{1}{k_0}\right)\right].
\end{equation}
For the reflected packet, let $t^{\prime}_1$ and
$t^{\prime}_2$ be such instances of time that

\begin{equation} \label{10}
<\hat{x}>_{inc}(t^{\prime}_1)
=<\hat{x}>_{ref}(t^{\prime}_2)=a-L_1.
\end{equation}

\noindent From equations (\ref{5}), (\ref{7}) and (\ref{10}) it
follows that the "reflection time" $\Delta t_{ref}$ ($\Delta
t_{ref}=t^{\prime}_2-t^{\prime}_1$) can be written as

\begin{equation} \label{11}
\Delta t_{ref}=\frac{m}{\hbar}\left[\frac{<J^\prime -
F^\prime>_{ref} +L_1} {<-k>_{ref}} +\frac{L_1}{k_0}+
a\left(\frac{1}{<-k>_{ref}} -\frac{1}{k_0}\right)\right].
\end{equation}

Notice that the expectation values of $k$ for all three wave packets
coincide only in the limit $l_0\to\infty$ (i.e., for particles with a
well-defined momentum). In the general case these quantities are
distinct (it is this asymptotic property of the wave packets that is
usually treated \cite{La1} as the acceleration (or retardation) of a
particle in tunneling). From Appendix (see (\ref{212})) it follows that
the rule

\begin{equation} \label{12}
\bar{T}\cdot <k>_{tr}+\bar{R}\cdot <-k>_{ref} =k_0
\end{equation}
must be true.

As is seen, times (\ref{9}) and (\ref{11}) cannot serve as
characteristic times for a particle.  Due to the last term in (\ref{9})
and (\ref{11}), the above times depend essentially on the initial
distance between the wave packet and barrier, with $L_1$ being fixed (a
similar peculiarity arises in \cite{Ja1}).  These contributions are
dominant for the sufficiently large distance $a$.  Moreover, one of
these terms, either in (\ref{9}) or in (\ref{11}), must be negative.
For example, it takes place for the transmitted wave packet in the case
of the under-barrier tunneling through an opaque rectangular barrier.
As was mentioned above, the numerical modelling of the tunneling
process (see, for example, \cite{Ha2, Ha1, Le1, Ter}) shows, in this
case, a premature appearance of the CM of the transmitted packet behind
of the barrier.  This fact have been treated in \cite{La1} as the
evidence of the lack of a causal link between the transmitted and
incident wave packets. To avoid this effect, the fields of application
of the SWPA have been restricted (see \cite{Ha2, Ha1}) by the limiting
case $l_0\to\infty$. A simple analysis shows however that the last
terms in (\ref{9}) and (\ref{11}) remain dominant in the limit $l_0\to
\infty$, with the ratio $l_0/a$ being fixed. Note that the limit
$l_0\to \infty$ with a fixed value of $a$ is unacceptable in this
analysis, because it contradicts the initial condition $a\gg l_0$ for a
completed scattering. Thus, even in the limit $l_0\to\infty$ the above
analysis carried out in the context of the SWPA does not provide the
well-defined characteristic times for a particle.

\section {Formalism of the individual description of the
sub- \\ ensembles of transmitted and reflected particles at the initial
stage of tunneling}

\hspace*{\parindent} We think that a principle mistake made in the
SWPA, in timing scattering particles, is that the "transmission time"
was obtained there from the analysis of the relative motion of the
transmitted wave packet with respect to the incident one. This is a
physically meaningless step. The point is that the incident packet
describes the whole quantum ensemble of particles, while the
transmitted packet presents only its part.  The incident packet can be
used as a reference only for the transmitted and reflected packets
taken jointly; in this case the initial and final states both describe
the whole ensemble of particles. In the individual description of
the transmitted (or reflected) packet, its motion should be compared
with the corresponding counterpart which describes the state of
transmitted (or reflected) particles at the initial stage of the
scattering event.  Searching for such counterparts for both
subensembles is the following step of our analysis.

\subsection*{Initial states of the subensembles of transmitted and
reflected particles}

Note that $k$-distributions for both subensembles at early times can be
obtained immediately if one takes into account the physical sense of the
transmission (and reflection) coefficient as well as the Born
interpretation of a wave function. So, the transmission coefficient
$T(k)$ is, by definition, the probability for the incident particle
with the momentum $\hbar k$ to pass ultimately through the barrier. The
expression $|f_{inc}(k,t)|^2 dk$ is the probability for the particle to
be, in $k$ space, at time $t$ in the interval $[k,k+dk]$.  Since both
the probabilities describe statistically independent events, the
expression $T(k)|f_{inc}(k,t)|^2 dk$ is evident to give the probability
that both these opportunities happen jointly for this particle (note
that this joint probability does not represent a conditional
probability introduced in \cite{Ste}; the former is always positive
unlike the latter). From this it follows that the function
$f_{inc}^{tr}(k,t)$, where

\begin{equation} \label{900}
f_{inc}^{tr}(k, t)=\sqrt{T(k)}\cdot
f_{inc}(k,t)\exp(i\theta(k)_{tr}),
\end{equation}
may be considered as a wave function describing at the initial stage
the subensemble of transmitted particles. Similarly, the function
$f_{inc}^{ref}(k,t),$ where
\begin{equation} \label{901}
f_{inc}^{ref}(k,t)=\sqrt{R(k)}\cdot
f_{inc}(k,t) \exp(i\theta(k)_{ref}),
\end{equation}
may be considered as a wave function to describe, at this stage,
reflected particles. Here $\theta_{tr}(k)$ and $\theta_{ref}(k)$ are
arbitrary real functions.

We see that the $k$-distributions for the to-be-transmitted and
transmitted wave packets are the same. The similar situation takes
place for reflected particles. The only difference is that the sign of
$k$ is different after the reflection. This property corresponds to
that of a classical scattering pointed out in \cite{Ja1} (see
Introduction).  To complete the correspondence, one needs to state the
initial condition for each scattering channel.  Namely, we will
consider that the CMs of the to-be-transmitted and to-be-reflected wave
packets should start at $t=0$ from the same point, i.e, from the
origin.

One can see that the above conditions are not sufficient to find
uniquely the phase shifts $\theta_{tr}(k)$ and $\theta_{ref}(k).$ It is
easily to show that there is a variety of wave packets, with the
different functions $\theta_{tr}(k)$ and $\theta_{ref}(k),$ which start
from the origin and described by the same $k$-distribution. However,
for the following, of great importance is only the fact that the
positions of CMs of all such packets coincide at any moment $t$.

For example, the initial states of the to-be-transmitted and
to-be-reflected subensembles, which obey the above requirements, can
be defined as follows

\begin{equation} \label{130}
\Psi_{inc}^{tr}(x)= \tilde{\Psi}_{inc}^{tr}(x-x_{tr}), \ppp
\Psi_{inc}^{ref}(x)=\tilde{\Psi}_{inc}^{ref}(x-x_{ref}),\ppp
\end{equation}
where

\begin{equation} \label{140}
\tilde{\Psi}_{inc}^{tr}(x)=\lim_{t\to \infty} e^{i\hat{H}_0 t/\hbar}
\hat{\Theta}_+  e^{-i\hat{H}t/\hbar}\Psi_0(x),\ppp
\tilde{\Psi}_{inc}^{ref}(x)=\lim_{t\to \infty} e^{i\hat{H}^{ref}_0
t/\hbar} \hat{\Theta}_- e^{-i\hat{H}t/\hbar}\Psi_0(x);
\end{equation}

\begin{equation} \label{150}
x_{tr}=\frac{<\tilde{\Psi}_{inc}^{tr}|\hat{x}|\tilde{\Psi}_{inc}^{tr}>}
{<\tilde{\Psi}_{inc}^{tr}|\tilde{\Psi}_{inc}^{tr}>};\ppp
x_{ref}=\frac{<\tilde{\Psi}_{inc}^{ref}|\hat{x}|
\tilde{\Psi}_{inc}^{ref}>}
{<\tilde{\Psi}_{inc}^{ref}|\tilde{\Psi}_{inc}^{ref}>}
\end{equation}

It is easy to show that the phases $\theta(k)_{tr}$ and
$\theta(k)_{ref}$ of the Fourier transforms $f_{inc}^{tr}(k,0)$ and
$f_{inc}^{ref}(k,0)$ of the initial states $\Psi_{inc}^{tr}(x)$ and
$\Psi_{inc}^{ref}(x)$ read as
\[\theta(k)_{tr}=J(k)-k<J'(k)>_{tr}\]
\[\theta(k)_{ref}=J(k)-F(k)-k<J'(k)-F'(k)>_{ref}.\]

Functions (\ref{130}) are evident to be non-orthogonal. In this case
the incident wave packet can be presented in the form

\begin{equation} \label{113}
f_{inc}(k,t)=f_{inc}^{tr}(k,t)+f_{inc}^{ref}(k,t)+
f_{inc}^{int}(k,t)
\end{equation}
\noindent where



\[f_{inc}^{int}=\left(1-\sqrt{T(k)}\cdot \exp(i\theta(k)_{tr})-
\sqrt{R(k)}\cdot \exp(i\theta(k)_{ref}\right)\cdot f_{inc}(k,t).\]

The non-orthogonality of the initial states of the transmitted and
reflected subensembles of particles is connected ultimately to the fact
that the future of a single particle from the incident quantum ensemble
is unpredictable in quantum mechanics. The sorting of particles at the
initial stage of scattering is, in fact, purely conditional. In other
words, there is an interchange of particles between the subensembles
corresponding to the to-be-transmitted and to-be-reflected wave packets.
Nevertheless, in spite of this exchange, the number of particles in
each subensemble remains constant. One can show that, in this case,
the following relations are valid,

\begin{equation} \label{16}
<f_{inc}^{tr}|f_{inc}^{tr}>+<f_{inc}^{ref}|f_{inc}^{ref}>
=<f_{inc}|f_{inc}>.
\end{equation}

\begin{equation} \label{17}
<f_{inc}^{tr}|f_{inc}^{tr}>=<f_{tr}|f_{tr}>,\ppp
<f_{inc}^{ref}|f_{inc}^{ref}>=<f_{ref}|f_{ref}>;
\end{equation}

\begin{equation} \label{18}
<k>_{inc}^{tr}=<k>_{tr};\ppp <k>_{inc}^{ref}=<-k>_{ref};
\end{equation}

\begin{equation} \label{19}
<\hat{x}>_{inc}^{tr}(t)=m^{-1}\hbar<k>_{inc}^{tr}\cdot t;
\end{equation}

\begin{equation} \label{20}
<\hat{x}>_{inc}^{ref}(t)=m^{-1}\hbar<k>_{inc}^{ref}\cdot t.
\end{equation}

As is seen, in (\ref{16}) there are neither terms with $f_{inc}^{int}$
nor terms to describe interference between the three contributions
entering decomposition (\ref{113}).  Besides, we have to emphasize
that, at this stage, there is no interference between the incident and
reflected wave packets (the latter has not yet been formed). Thus, from
these expressions it follows that, long before the collision, the whole
quantum ensemble of incident particles do indeed consists from two
parts (see (\ref{16})).  Relations (\ref{17}) suggest that, at this
stage, the to-be-transmitted and to-be-reflected parts are described
by $f_{inc}^{tr}(k,t)$ and $f_{inc}^{ref}(k,t),$ respectively. For
each subensemble, both the number of particles (see (\ref{17})) and the
absolute value of the average wave-number $k$ (see (\ref{18})) must
be the same both before and after the collision. One can easily show
also that

\begin{equation} \label{21}
\bar{T}\cdot<k>_{inc}^{tr}+ \bar{R}\cdot<k>_{inc}^{ref}=k_0.
\end{equation}

\noindent Besides, at $t=0$ the CMs of both the packets are located, in
accordance with the initial conditions, at the point $x=0$ (see
(\ref{19}) and (\ref{20})).

Thus, at the stage long before the scattering event, the incident wave
packet can be divided into the to-be-transmitted and to-be-reflected
ones whose evolution is described uniquely by relations (\ref{16}) -
(\ref{21}). This provides solid grounds to say now that tunneling is a
two-channel scattering, and (answering the questions in \cite{La1}),
that the potential barrier does not accelerate (on the average) a
particle. By our approach, the absolute value of the average momentum
of particles must be conserved asymptotically not only for the whole
quantum ensemble, but also for its transmitted and reflected parts. We
have to stress that this conservation law was stated in the frame of
conventional quantum mechanics, without new guesses.

\subsection*{On the experimental verification of the individual
description of scattering channels in tunneling}

At first sight, it seems to be strange that, although the need for an
individual description of both scattering channels has been recognized
(see Introduction), a desirable formalism have not yet been elaborated.
In our opinion, the main obstacle which arises here is the problem of
the interpretation of quantum mechanics, rather than a purely
mathematical one. For example, according to the orthodox
interpretation, the above formalism contradicts the causality
principle since the knowledge of the initial states for both
scattering channels should mean that a particle "feels" the barrier yet
before the interaction (see, for example, \cite{Ja1}). We disagree
entirely with such a viewpoint.  It itself contradicts Born's
interpretation of a wave function. We think that a proper understanding
of the above formalism can be attained only in the framework of the SI
which is known to base entirely on the statistical, Born interpretation
of a wave function. According to the SI, our formalism does not
contradict the causality principle, and can be verified experimentally.
Let us consider the last question in detail.

In order to verify our formalism, it is sufficient to check the fact
that the ensemble of incident particles can indeed be divided into two
parts, with the expectation values of the particle's position changing
in accordance with relations (\ref{19}) and (\ref{20}). To perform this
check, one needs to obtain the momentum and position (further, in this
section, these variables will be denoted by $P$ and $Q$, respectively)
distributions for the ensemble of incident particles. For the
subensembles of transmitted and reflected particles, one needs to
obtain only the $P$-distributions. All the above suggests carrying out
the sufficiently large (strictly speaking, infinite) number of
identical experiments to provide the needed $P$- and $Q$-data. In order
to satisfy conditions (\ref{70}) - (\ref{72}) all measurements must be
performed far enough from the barrier.

The basic asymptotic property of the subensembles which can be used in
this testing is that $P$-distributions for each scattering channel,
before and after the scattering event, must be the same. Thus, one
needs firstly to obtain experimental $P$-data for the subensembles of
transmitted and reflected particles, to extract from them the
corresponding $P$-distributions, and then to sort out incident particles
into two subsets with the given $P$-distributions. The last step is the
most important in this testing. Its peculiarity is that sorting the
$P$-data should give two corresponding subsets of the $Q$-data. This
means that the above $P$- and $Q$-data must represent the set of
$(P,Q)$-pairs. It is naturally to suppose that such pairs can be
obtained as a result of a simultaneous measuring of the particle's
position and momentum. According to the SI, "...there is no conflict
with quantum theory in thinking of a particle as having definite
[unpredictable] values of both position and momentum..." \cite{Bal}.
Moreover, we have to add that the values of $P$ and $Q$ can be
measured simultaneously (jointly) with an arbitrary accuracy (in
our opinion, if one assumed that the Heisenberg uncertainty relation
does not admit an accurate simultaneous measuring of these quantities,
he would then to recognize that this relation cannot be verified in
quantum mechanics). We consider that, strictly speaking, there are no
restrictions, in the conventional quantum mechanics, to measure jointly
these quantities. In particular, this means that one can construct,
in principle, such devices that will allow one to measure $P$ and $Q$
simultaneously, without the violation of the connection between the
$P$- and $Q$-data imposed by the Fourier-transformation. Thus, in line
with the terminology introduced in \cite{Bus}, in this question we
adhere the second extreme view.  In any case, the possibility to
measure jointly $P$ and $Q$ have been accepted, in principle, by many
authors (see, for example, \cite{Bal, Art, Mu5, Bra}).

So, let for the given instant $t$ $(P,Q)$-pairs for incident
particles, and $P$-data for scattered particles be available: it is
supposed that the total number of the experimental data is sufficiently
large and the $(P,Q)$-pairs have been accidentally numerated. So, one
may consider the $P$-distributions for the transmitted and reflected
subensembles have been known. Let also the $P$-scale be partitioned
into sets of intervals with a sufficiently small width $\hbar\Delta k$.
Then, let us consider for incident particles the $(P,Q)$-pairs with the
$P$ values to belong some interval $[\hbar k,\hbar(k+\Delta k)]$, take
from this interval the $T(k)$-th part of these pairs (for example, with
the first numbers) and include this part into the to-be-transmitted
subensemble.  Note that the values of $Q$ must not be taken into
account in sorting. Another part should be associated to the
to-be-reflected subensemble. All the above should be done for all
$\hbar k$-intervals.  Then, for incident particles of both scattering
channels, we can calculate the expectation values of $P$ and $Q$, for
the given $t$, and check lastly expressions (\ref{16})-(\ref{20}). Of
course, in doing so, we wait that the $P$- and $Q$-distributions for
each subensemble are connected by the Fourier-transformation.
Otherwise, they do not correspond some wave functions.

For the following it is important to consider another variant of
testing. Namely, let all measurements be carried out at a given point
$Z_1$ rather than at a given instant of time. Now we will measure the
particle's momentum P and the value of parameter $t$ at which the
incident particle arrives at the given point $Z_1$. Since before the
interaction the incident particle is free, the $P$-shape of the
corresponding incident wave packet must remain unaltered in time. This
means that such measurements should give the same $P$-distribution for
the incident packet, as in the above case. Thus, the only difference
between both these variants is that now one needs to sort $(P,t)$-pairs
instead of $(P,Q)$-pairs.  After the mean values of $t$ for each
subensemble have been calculated expressions (\ref{16})-(\ref{20}) can
be checked.  This variant is more suitable for testing characteristic
times introduced below.

\section {Delay times for the subensembles of transmitted and reflected
particles}

\newcommand {\uta} {\tau _ {tr}}
\newcommand {\utb} {\tau _ {ref}}
\hspace*{\parindent} Now, for both scattering channels, we can offer
new definitions of characteristic times to describe the influence of
the barrier on a particle. For this purpose we have again to consider
the particle's motion in a wide spatial interval containing the
barrier region. In particular, let us calculate the (average)
transmission time, $\uta$, spent by a particle in the interval $[Z_1,
Z_2]$. We have to take into account the fact that before the
interaction the motion of the to-be-transmitted wave packet is
described by the function $e^{-i\hat{H}_0t/ \hbar}\Psi_{inc}^{tr}(x)$
with the Fourier transform $f_{inc}^{tr}(k,t)$ (see expression
(\ref{900}), (\ref{130}) - (\ref{150})). Then the time-of-arrivals
$t_1$ and $t_2$ to correspond the extreme points $Z_1$ and $Z_2$,
respectively, must obey the equations

\begin{equation} \label{22}
<\hat{x}>_{inc}^{tr}(t^{tr}_1)=a-L_1; \ppp
<\hat{x}>_{tr}(t^{tr}_2)=b+L_2.
\end{equation}

\noindent From this it follows that

\begin{equation} \label{23}
\uta(L_1,L_2)\equiv t^{tr}_2-t^{tr}_1 =\frac{m}{\hbar
<k>_{tr}}\left(<J^\prime>_{tr} +L_1+L_2 \right).
\end{equation}

Similarly, let the reflection time, $\utb^{(-)}$, be the difference
$t^{ref}_2-t^{ref}_1$ where

\begin{equation} \label{24}
<\hat{x}>_{inc}^{ref}(t^{ref}_1)=<\hat{x}>_{ref}(t_2^{ref})=a-L_1,
\end{equation}
(remind that the to-be-reflected packet is described at the
initial stage of scattering by $e^{-i\hat{H}_0t/\hbar}
\Psi_{inc}^{ref}(x)$ with the Fourier transform $f_{inc}^{ref}(k,t)$
(see expression (\ref{901}), (\ref{130}) - (\ref{150}))

\noindent One can easily show that

\begin{equation} \label{25}
\utb^{(-)}(L_1)=\frac{m}{\hbar <-k>_{ref}}\left(<J^\prime -
F^\prime>_{ref} +2L_1\right).
\end{equation}

As was shown in \cite{Ch1, Ch2}, the sign of the phase $F^\prime$ is
opposite for waves impinging the barrier from the right. This case is
similar to that when the wave moves from the left to the potential
barrier $V_{inv}(x)$ such that $V_{inv}(x)=V(a+b-x)$. The corresponding
reflection time $\utb^{(+)}$ can be written as

\begin{equation} \label{26}
\utb^{(+)}(L_1)=\frac{m}{\hbar <-k>_{ref}}\left(<J^\prime +
F^\prime>_{ref} +2L_1\right).
\end{equation}

One has to stress once more that these quantities cannot be treated, at
$L_1=L_2=0$, as the transmission and reflection times for the barrier
region. There is no experiment to measure their values in this case, as
the disposition of devices at the boundaries of the barrier does not
provide a reliable identification of transmitted and reflected
particles: conditions (\ref{70}) - (\ref{72}) are not fulfilled in this
case.

Note that times (\ref{24}) - (\ref{26}) are not suitable to describe
the influence of the barrier on a particle, because they include the
contributions of the out-of-barrier regions. However, they enables one
to define time delays to characterize the relative motion of a
scattered and corresponding free particles. To determine these
quantities, we have to take into account the fact that beyond the
scattering region the mean velocity of a free particle (taken as a
reference for each scattering channel) should coincide with that of a
scattered particle. Then for transmitted particles the delay time
$\tau^{tr}_{del}$ can be written, in terms of wave functions (\ref{140})
and the tunneling parameters, as

\begin{equation} \label{27}
\tau^{tr}_{del}=\frac{m} {\hbar<k>_{tr}}\left(<J^\prime>_{tr} -d \right)
=\frac{<\tilde{\Psi}_{inc}^{tr}|\hat{x}|\tilde{\Psi}_{inc}^{tr}>}
{<\tilde{\Psi}_{inc}^{tr}|\tilde{\Psi}_{inc}^{tr}>}\left/
\frac{<\tilde{\Psi}_{inc}^{tr}|\hat{p}/m|\tilde{\Psi}_{inc}^{tr}>}
{<\tilde{\Psi}_{inc}^{tr}|\tilde{\Psi}_{inc}^{tr}>}\right. .
\end{equation}
Similarly, for reflected particles the delay time $\tau^{(-)}_{del}$
can be written as

\begin{equation} \label{28}
\tau^{(-)}_{del}=\frac{m}{\hbar <-k>_{ref}}\left(<J^\prime -
F^\prime>_{ref} -d\right) =
\frac{<\tilde{\Psi}_{inc}^{ref}|\hat{x}|\tilde{\Psi}_{inc}^{ref}>}
{<\tilde{\Psi}_{inc}^{ref}|\tilde{\Psi}_{inc}^{ref}>}\left/
\frac{<\tilde{\Psi}_{inc}^{ref}|\hat{p}/m|\tilde{\Psi}_{inc}^{ref}>}
{<\tilde{\Psi}_{inc}^{ref}|\tilde{\Psi}_{inc}^{ref}>}\right. .
\end{equation}
The delay time for the corresponding inverted barrier $V_{inv}(x)$ is
given by

\begin{equation} \label{29}
\tau^{(+)}_{del}=\frac{m}{\hbar <-k>_{ref}}\left(<J^\prime +
F^\prime>_{ref} -d\right)
\end{equation}
Expressions (\ref{27}) and (\ref{28}) (or (\ref{29}) for the inverted
barrier) should be considered as the definitions of delay times in our
approach. As is seen, these times do not depend on the initial distance
between the incident packet and barrier. Besides, they do not equal to
zero for the $\delta$-potential, since $<J^\prime>_{tr}$ and $<J^\prime
+ F^\prime>_{ref}$ do not vanish in this case.

It is evident that $x_{tr}$ (see expressions (\ref{150})) can be
treated as the spatial delay for the subensemble of transmitted
particles, with $x_{tr}=<J^\prime>_{tr} -d$. Similarly, $x_{ref}$
with $x_{ref}=<J^\prime - F^\prime>_{ref} -d$ can be treated as the
spatial delay for reflected particles. In the case of the inverted
barrier, the spatial delay is equal to $<J^\prime + F^\prime>_{ref}
-d$.

It is useful to compare respectively expressions (\ref{23}) and
(\ref{25}) with (4.8) and (4.9) presented in \cite{Ha2}. Unlike our
definitions, the integral in expression (4.8) is divergent at the
point $k=0$ if $A(0;k_0,l_0)\ne 0$ and $T(0)\ne 0$. For expression
(4.9) a similar situation arises if $A(0;k_0,l_0)\ne 0$ and $R(0)\ne
0$. See also \cite{Nus, Mu3}.

\section {The low bound of the scattering time}

It should be noted that the above delays for both subensembles are
accumulated during the interaction of a particle with the barrier. In
this case the delay times themselves say nothing about the duration of
this process. Thus, one has to define the third characteristic time of
tunneling, which will be further referred to as the low bound of the
scattering time.  It is obvious that this quantity cannot be defined
individually for each scattering channel, because it should describe
just the very stage of the one-dimensional scattering when a particle
cannot be identified as incident, transmitted or reflected one.
Besides, in this case one should keep in mind that the scattering event
lasts until the probability to find a particle in the barrier region is
noticeable.  This means that in defining this quantity, one should to
take into account spreading the corresponding wave packet. It is
obvious that the scattering time does not coincide with the time of
staying the particle in the barrier region (see Introduction).
Formally speaking, to define the scattering time, one needs to find
such temporal interval for which conditions (\ref{70}) - (\ref{72}) are
violated.

So, let us define such instant of time, $t_{start}$, at which the
distance between the CM of the incident packet and the left boundary of
the barrier is equal to the half-width of this packet, i.e.,

\begin{equation} \label{300}
(a-<\hat{x}>_{inc}(t_{start}))^2=<(\delta \hat{x})^2>_{inc}(t_{start}).
\end{equation}
We will treat the instant $t_{start}$ as the time of starting the
scattering event. We will assume that for $t\le t_{start}$ conditions
(\ref{70}) - (\ref{72}) are fulfilled yet with a sufficient accuracy.

As regards the end of the scattering event, one has to take into
account that the transmitted and reflected particles move, on this
stage, in the disjoint spatial regions, i.e., their wave packets do not
interfere with each other. To define the corresponding instant of time
$t_{end}$, one has to use the quantities $S_{tr+ref}$ and
$<(\delta\hat{x})^2>_{tr+ref}$ characterizing jointly both these packets
(see (\ref{1190}) and (\ref{401}) in Appendix). Here $S_{tr+ref}$ is a
total mean distance between the CMs of the transmitted and reflected
packets, and the corresponding nearest boundaries of the barrier;
$<(\delta\hat{x})^2>_{tr+ref}$ is a total mean-square deviation of
$\hat{x}$ averaged over the transmitted and reflected packets.

Thus, let $t_{end}$ be the instant of time at which

\begin{equation} \label{301}
S^2_{tr+ref}(t_{end})=<(\delta\hat{x})^2>_{tr+ref}(t_{end}).
\end{equation}
This definition suggests that for $t\ge t_{end}$ conditions (\ref{70}) -
(\ref{72}) are fulfilled with a sufficient accuracy.

Either of the above equations has two roots. A simple analysis shows
that one should take the smallest root, in the case of equation
(\ref{300}), and the biggest root, in the case of (\ref{301}). So, the
searched-for solutions to (\ref{300}) and (\ref{301}) can be written in
the form

\begin{equation} \label{302}
t_{start}=\frac{m}{\hbar}\cdot\frac{a k_0-\sqrt{l^2_0k_0^2
+(a^2-l_0^2)<(\delta k)^2>_{inc}}}{k_0^2 - <(\delta k)^2>_{inc}};
\end{equation}
(remind that $a\gg l_0$);

\begin{equation} \label{303}
t_{end}=\frac{m}{\hbar} \cdot \frac{\bar{b}
k_0-\chi+ \sqrt{l^2k_0^2+\chi^2-2k_0\bar{b}\chi+
(\bar{b}^2-l^2)<(\delta k)^2>_{tr+ref}}} {k_0^2 -<(\delta
k)^2>_{tr+ref}}
\end{equation}
(see (\ref{224})-(\ref{227})). The low bound of the scattering time
$\tau_{scatt}$ can be defined now as the difference
$t_{end}-t_{start}$.

A simple analysis shows that this quantity is strictly positive when
the inequality

\begin{equation} \label{304}
(k_0^2-<(\delta k)^2>_{inc}) > 0
\end{equation}
is fulfilled. It should be considered as a condition for a completed
scattering. It guarantees that (almost) all incident particles start
at $t=0$ toward the barrier, and that the transmitted and reflected
packets occupy, in the limit $t\to\infty$, the disjoint spatial
regions.  For a completed scattering the hierarchy
$t_{end}>t_{start}>0$ is obvious to take place. The second inequality
is guaranteed by (\ref{304}). The first one must be valid, since the
number of particles in the whole quantum ensemble is constant. Or, in
other words, the quantum ensemble of particles cannot exit a region
before entering it. In this case it is important to note that, in the
limit $k_0^2 \to <(\delta k)^2>_{inc}$,
\[t_{start}=\frac{m(a^2-l_0^2)}{2\hbar k_0 a}\approx \frac{m a}{2\hbar
k_0}.\]  That is, expression (\ref{302}) has no singularity in this
limit.

One can easily show that there is the optimal value of $l_0$ at which
$\tau_{scatt}$ is minimal. The point is that, in the limit $l_0\to
\infty$, the scattering time grows together with $l_0$, and, at small
values of this parameter, this time is large because of the fast
spreading of the wave packet. If requirement (\ref {304}) is violated,
the transmitted and reflected packets must be overlapped at
$t\to\infty$ due to their spreading. As a result, the scattering event
becomes incomplete. In this case the process of scattering a particle
mimics that of escaping the particle from the barrier region. Our
approach allows one to define the scattering time for the incomplete
scattering too. However this question is beyond the framework of our
paper.

In the limit $l_0\to\infty$, expressions (\ref{302}) and (\ref{303})
are esensially simplified. In this case we have $l\approx l_0,$
$k_0^2\gg <(\delta k)^2>_{inc}.$ Thus, the spreading effect may be
neglected for narrow (in $k$-space) wave packets. Taking account only
of the dominant terms in (\ref{302}) and (\ref{303}), we obtain

\begin{equation} \label{32}
\tau_{scatt}=\frac{m}{\hbar k_0}\left(2l_0+<J^\prime>_{inc} - <R
F^\prime>_{inc} \right).
\end{equation}
For the corresponding inverted barrier the last term in (\ref{32}) has
an opposite sign. This term is nonzero only for asymmetrical potential
barriers.

One has to stress once again that the above definition yields the
low bound of the scattering time. For it does not take into account the
"tails" of wave packets. This means that, if $\bar{T}\ll 1$ (or
$\bar{R}\ll 1$), the to-be-transmitted (or to-be-reflected) wave packet
may occur in the scattering region when $t\not\in [t_{start},
t_{end}]$. To correct $\tau_{scatt}$ for these cases, one needs to
"reduce" the size of the packet's tails. Namely, for these cases a part
of particles in the "tails" must be smaller than min($\bar{T},
\bar{R}$). It is obvious that such a correction should increase the
scattering time. However, we think this is superfluous. The point is
that the quantity $\tau_{scatt}$ defined above is sufficient to
estimate the duration of the scattering process for the body of the
quantum ensemble of particles.  As regards its small part (transmitted
or reflected), the instants of time $t^{tr}_1$ and $t^{tr}_2$ (or
$t^{ref}_1$ and $t^{ref}_2$) given by equations (\ref{22}) (or
(\ref{24})) provide a sufficient information about its motion.

\section{Tunneling the Gaussian wave packet through rec\-tangular
barriers}

To display some properties of tunneling, we have considered rectangular
potential barriers, and investigated in detail the tunneling parameters
of a particle whose initial state is described by the Gaussian wave
packet (GWP). The weight function $A\aro$ in (\ref{3}) is defined in
this case by the expression

\[A\aro=\exp(-l_0^2(k-k_0)^2).\]

As was pointed out above, in the general case the asymptotic value of
the average wave-number of transmitted particles differs from that of
reflected particles. One can show that for the GWP

\begin{equation} \label{100}
<k>_{tr}=k_0+\frac{<T^\prime>_{inc}} {4l_0^2<T>_{inc}};
\end{equation}

\begin{equation} \label{101}
<-k>_{ref}=k_0+\frac{<R^\prime>_{inc}} {4l_0^2<R>_{inc}}.
\end{equation}
Supposing that

\[<k>_{tr}=k_0+(\Delta k)_{tr},\ppp <-k>_{ref}=k_0+(\Delta k)_{ref},\]

\noindent we can rewrite relations (\ref{100}) and (\ref{101}) in the
form

\begin{equation} \label{102}
\bar{T}\cdot (\Delta k)_{tr}=-\bar{R}\cdot(\Delta k)_{ref}
=\frac{<T^\prime>_{inc}}{4l_0^2}.
\end{equation}
\noindent Note that $R^\prime=-T^\prime$. The first equation in
(\ref{102}) coincides with rule (\ref{12}) to be valid for any wave
packet.

Let us also derive several useful correlations for the mean-square
deviations of the $\hat{k}$ operator. Using the relations (\ref{511})
in Appendix, one can show that for the GWP

\[<(\delta k)^2>_{inc}=\frac{1}{4l_0^2};\]

\[<(\delta k)^2>_{tr}=\frac{<T(k)(k-k_0)^2>_{inc}}{<T>_{inc}}
-\frac{(\Delta k)_{tr}}{2l_0^2} \frac{<T'>_{inc}}{<T>_{inc}} +(\Delta
k)_{tr}^2;\]

\[<(\delta k)^2>_{ref}=\frac{<R(k)(k-k_0)^2>_{inc}}{<R>_{inc}}
-\frac{(\Delta k)_{ref}}{2l_0^2} \frac{<R'>_{inc}}{<R>_{inc}} +(\Delta
k)_{ref}^2.\]
Further calculations (see (\ref{226})) yield

\begin{equation} \label{500}
<(\delta k)^2>_{tr+ref}=\frac{1}{4l_0^2} \left(1-\frac{<T'>_{inc}^2}
{4l_0^2\bar{T}\bar{R}}\right).
\end{equation}

Now we dwell shortly on the numerical analysis of some peculiarities
of tunneling (a more detailed numerical analysis is supposed to be
done in the following).  All calculations have been carried out on
the basis of the TMM \cite{Ch1} (see also \cite{Ch3}). To clear up the
role of the spatial localization of a tunneling particle, we have
investigated the main features of the $\l_0$-dependence of the
tunneling parameters for the particular cases when $V(x)=V_0=0.3 eV$,
$E_0=0.02 eV$; $m= 0.067m_e$ where $m_e$ is the electron mass.
Fig.1 shows the ratios of the mean wave numbers of the transmitted
and reflected packets to that of the incident packet versus
$\log(d/l_0)$.  As is seen, for particles with the well-defined
momentum or position, the expectation values of the momentum for the
transmitted and incident packets are the same. The reason is that in
the case of the well-defined momentum the incident packet consists,
in fact, of a single wave. In the second case the particle tunnels
through the barrier without reflection. The point is that the
contribution of high-energy harmonics into the GWP, for which the
barrier is more transparent, grows together with $d/l_0$.  As a
result, the transmission coefficient increases too (see Fig. 1).  In
the domain $0\leq\log(d/l_0)\leq 2$ a situation arises when the
contributions of the transmitted and reflected waves are approximately
equal. In this case it takes place the most distortion of the packet's
shape as well as the maximal difference between the expectation values
of the momentum for the transmitted and incident wave packet. In the
case of a many-barrier structure, noticeable variations of
$\overline{T}(d/l_0)$ can be observed if there are single-wave
resonances near the point $E_0$ ($E_0=E(k_0)$).  However, in any case,
$\overline{T}\to 1$, in the limit $l_0\to 0$, irrespective of the
barrier's shape and value of $k_0$.

Now we address to the spatial delays $\langle J^\prime-d\rangle_{tr}$
and $\langle J^\prime-d\rangle_{ref}$. As was shown in \cite{Ch2}, for
the rectangular barrier of height $V_0$ and width $d$, $F^\prime
\equiv 0$ and the derivative $J^\prime$ is determined by the expressions

\begin{equation} \label{103}
J^\prime= \frac{2(\kappa^2-k^2)k^2\kappa d
+(k^2+\kappa^2)^2\sinh(2\kappa d)}{\kappa[4k^2\kappa^2+
(k^2+\kappa^2)^2\sinh^2(\kappa d)]},
\end{equation}
where
$\kappa=\sqrt{2m(V_0-E)/\hbar^2},$ $E<V_0;$

\begin{equation} \label{104}
J^\prime= \frac{2(\kappa^2+k^2)k^2\kappa d
-(k^2-\kappa^2)^2\sin(2\kappa d)}{\kappa[4k^2\kappa^2+
(k^2-\kappa^2)^2\sin^2(\kappa d)]},
\end{equation}
where
$\kappa=\sqrt{2m(E-V_0)/\hbar^2},$ $E\ge V_0.$

This enables one to explain the numerical data obtained for $\langle
J^\prime-d\rangle_{tr}$ and $\langle J^\prime- d \rangle_{ref}$.  As is
seen from Fig. 2, both quantities are equal only for particles with the
well-defined momentum. It is important that in the limit $l_0\to 0$ the
(average) spatial delay for a transmitted particle vanishes. This
property, taking place for any barrier, is due to the fact that the
average energy of a particle grows infinitely in this limit (such a
particle passes freely through the barrier). As regards reflected
particles, for rectangular barriers the spatial delay tends to
$[J^\prime(k)-d]|_{k=0}$.

It is interesting also to consider the case of the under-barrier
tunneling, providing that the barrier's width grows but the wave-packet
width is fixed. It is the very case which is usually analyzed in
the literature (e.g., see \cite{Ha2, Ja1, Kar, Yam}) to demonstrate the
"superluminal" propagation of particles in tunneling. From (\ref{103})
it follows that for $E<V_0$ and $\kappa d\gg 1,$

\begin{equation} \label{163}
J^\prime \approx 2\kappa^{-1},
\end{equation}
i.e., in this case, for a particle with a well-defined momentum, this
quantity does not depend on $d$. For $E>V_0$ and $\kappa d\gg 1,$
\begin{equation} \label{164}
J^\prime \approx
\frac{2(\kappa^2+k^2) k^2 d}{4k^2\kappa^2+ k_{top}^4 \sin^2(\kappa d)}
\end{equation}
Besides, for $E=V_0$
\begin{equation} \label{165}
J^\prime = \frac{2}{3}\cdot \frac{9+2k_{top}^2 d^2}{4+k_{top}^2 d^2}
\cdot d
\end{equation}
where $k_{top}=\sqrt{2mV_0/\hbar^2}$. From (\ref{163}) - (\ref{165}) it
follows that for large enough $d$ the function $J^\prime(k)$ is
smooth for $E<V_0$ and rapidly oscillates for $E>V_0$. Besides, it
grows sharply in going from the "under-barrier" domain to the
"above-barrier" one. A simple analysis shows that the wider the
barrier, the larger the contribution of waves with $E>V_0$ to the
transmitted packet. In this case the transmission coefficient decreases
until waves with $E<V_0$ dominate in the transmitted wave packet.
However, when these waves have been filtered off this packet,
$\overline{T}$ and $<k>_{tr}$ does not depend on $d$ (see Fig. 3 and
Fig. 4). These quantities, as functions of $d$ which was varied from
5 nm to 200 nm, have been calculated for $l_0=15$ nm; other parameters
are the same as for fig.1.  As is seen, for the sufficiently large $d$,
$<k>_{tr}\approx k_{top}.$ That is, the average energy of particles
transmitted through a wide opaque rectangular barrier exceeds slightly
the barrier height. Everywhere in the $d$-interval investigated the
average energy of reflected particles coincides approximately with that
of incident particles; the point is that here $\overline{R}\approx 1$.

Besides, we found (see Fig. 5) that in the "under-barrier" domain
$<J^\prime>_{tr} \approx <J^\prime>_{ref} \approx 2.86$ nm. In the
"above-barrier" domain, the behavior of $<J^\prime>_{tr}(d)$ is
described qualitatively by expression (\ref{164}).  As regards
reflected particles, $<J^\prime>_{ref}$ amounts about $2.86$ nm in both
these domains.  For the case considered, $\tau_{scatt}\approx 0.143$ ps.

\section*{Conclusion}

In this paper we have developed a new variant of the wave-packet
analysis and proposed a solution to the tunneling time problem. Its key
point is a formalism of the individual asymptotic description of
transmitted and reflected particles at the stage preceding the
scattering event. We have shown that this formalism does not contradict
the principles of quantum mechanics, and it can be verified
experimentally. According to our approach, the tunneling process, being
an elastic one-dimensional scattering of one particle on the
time-independent potential barrier, must be treated as a two-channel
scattering. For each scattering channel, the momentum distribution
must conserve asymptotically.

It is shown that the above formalism gives a good grounds to solve
the tunneling time problem. For each scattering channel we have defined
delay times which describe the relative motion of the scattered and
corresponding free particle moving, outside the scattering region, with
the same average velocity. In addition, to characterize the duration of
the interaction of a particle with the potential barrier, we define the
low bound of the scattering time. It describes jointly both scattering
channels. As is shown, there is an optimum initial width of the
incident wave packet, leading to the minimal value of the scattering
time. This time is strictly positive and describes the main part of
particles of the quantum ensemble.

All three characteristic times are defined for wave packets of an
arbitrary width. They are obtained in terms of the expectation values
of the position and momentum operators (in this sense, characteristic
times in the given approach are quantities of a secondary importance).
It is important that all integrals arising in our formalism have no
singularity at $k=0$, however the transmission coefficient and function
$A\aro$ depend on $k$ in the vicinity of $k=0$. This property may be
useful, in particular, in studying the "ultra Hartmann effect" (see
\cite{Mu3}).

As is shown, the scattering process is completed, if only the CM
of the incident wave packet moves quickly enough (see condition
(\ref{304}). Otherwise, the problem of scattering a particle off the
potential barrier turns, in fact, into that of escaping the particle
from the barrier region. Note that it is quite possible to adapt our
formalism to this case too.

\section*{Appendix: the asymptotic properties of a wave func\-tion in
the $k$-representation}

The peculiarity of a completed scattering is that at the initial
instant of time the incident packet is located entirely to the left of
the barrier. After the scattering event there is two packets moving
away from the barrier. In the limit $t\to \infty$, they are located in
the disjoint spatial regions.

It is suitable to present the wave functions describing the incident,
transmitted and reflected packets in the form

\newcommand {\intk}{\int_{-\infty}^{\infty}dk}
\newcommand {\intx}{\int_{-\infty}^{\infty}dx}
\newcommand {\da}{\partial}

\begin{equation} \label{201}
\Psi(x,t)=\frac{1}{\sqrt{2\pi}} \intk f(k,t) e^{ikx},
\end{equation}
where $f(k,t)\in {\cal S}_{\infty};$
\[f(k,t)=M\aro \exp(i\xi(k,t));\]
$M\aro$ and $\xi(k,t)$ are the real functions. In particular, for the
incident packet

\begin{equation} \label{202}
M_{inc}\aro=c A\aro; \ppp
\xi_{inc}(k,t)=-\frac{\hbar k^2 t}{2m}.
\end{equation}
For the transmitted and reflected packets we have
\begin{equation} \label{203}
M_{tr}\aro=\sqrt{T(k)}M_{inc}\aro;  \mbox{\hspace{3mm}}
\xi_{tr}(k,t)=\xi_{inc}(k,t)+J(k)-kd;
\end{equation}
\begin{equation} \label{204}
M_{ref}(-k,k_0)=\sqrt{R(k)}M_{inc}\aro; \mbox{\hspace{2mm}}
\xi_{ref}(-k,t)=\xi_{inc}(k,t)+2ka+J(k)-F(k)-\frac{\pi}{2}.
\end{equation}
In the case of a completed scattering the above packets provide the
asymptotic behavior of a wave function.

For any Hermitian operator $\hat{Q}$, Fourier transformation
(\ref{201})-(\ref{204}) enables one to determine the evolution of the
expectation value $<\hat{Q}>$,
\begin{equation} \label{205}
<\hat{Q}>=\frac{<\Psi|\hat{Q}|\Psi>}{<\Psi|\Psi>},
\end{equation}
at the stages preceding and following the scattering event, where $\Psi$
is one of the above wave packets. That is, both for the whole quantum
ensemble and for its subensembles, we will calculate the expectation
values of $\hat{Q}$ by means of the same rule (\ref{205}).

Strictly speaking, for the incident and reflected packets, the
integrals in (\ref{205}) should be calculated over the interval
$(-\infty, a]$. For the transmitted packet, one should integrate over
the interval $[b,\infty)$. The point is that expressions (\ref{201})
-(\ref{204}) for these packets are valid only for the corresponding
spatial region and corresponding stage of scattering. However, taking
into account that the body of each packet is located, in the limit
$t\to \infty$ or $t\to -\infty$, in its "own" spatial region, we may
extend the integration in (\ref{205}) onto the whole $OX$-axis.  Due to
this step the description of these packets becomes very simple.  At the
same time, the mistake introduced in the formalism is expected to be
negligible: the farther is the packet from the barrier at the initial
time, the smaller is this mistake. It vanishes in the limits $t\to
\infty$. Thus, the asymptotic wave function for a completed scattering
(for $\Psi_0\in S_{\infty}$) may be studied in the $k$-representation.
Now, on the basis of this representation, we can find the main
characteristics of all three packets, which are desirable for the
following.

\subsection*{Normalization}

Note that
\[<\Psi|\Psi>=\frac{1}{2\pi} \intx \intk dk'
f^*(k',t)f(k,t)\exp[i(k-k')x] =\]
\begin{equation} \label{206}
=\intk |f(k,t)|^2= \intk M^2\aro.
\end{equation}
For each packet we have then the following norms. Since the particle is
located at the initial time to the left of the barrier, we have

\begin{equation} \label{207}
<\Psi_0|\Psi_0>=\intk M^2_{inc}\aro=1.
\end{equation}
Then, allowing for (\ref{202}), we have
\begin{equation} \label{208}
c^{-2}=\intk A^2\aro.
\end{equation}
For the transmitted packet,
\begin{equation} \label{209}
<f|f>_{tr}=\intk M^2_{tr}\aro =\intk T(k)M^2_{inc}\aro\equiv
<T(k)>_{inc}\equiv \bar{T}.
\end{equation}
For the reflected packet,
\[<f|f>_{ref}=\intk M^2_{ref}\aro =\intk R(k)M^2_{inc}(-k;k_0).\]
Having made an obvious change of variables, we obtain
\begin{equation} \label{210}
<f|f>_{ref}=<R(k)>_{inc}\equiv \bar{R}.
\end{equation}
From (\ref{207}) - (\ref{210}) it follows that
\[\bar{T}+\bar{R}=1.\]

\subsection*{The expectation values of the operators $\hat{k}^n$ ($n$
is the positive number)}

Considering (\ref{201}), one can find for all the packets that
\[<\Psi|\hat{k}|\Psi>=-i<\Psi|\frac{\da\Psi}{\da x}>=\intk M^2\aro k\]
(i.e., $\hat{k}$ is a multiplication operator in this case).
Then for any value of $n$ we have
\begin{equation} \label{211}
<\Psi|\hat{k}^n|\Psi>=<f|k^n|f>=\intk M^2\aro k^n.
\end{equation}
Now we can treat the individual packets. From (\ref{211}) and
(\ref{203}) it follows that
\[<f_{tr}|k^n|f_{tr}>=<f_{inc}|T(k)k^n|f_{inc}>.\]
In a similar way we find also that
\[<f_{ref}|k^n|f_{ref}>=(-1)^n<f_{inc}|R(k)k^n|f_{inc}>,\]
and, hence,
\begin{equation} \label{511}
<T(k)k^n>_{inc}=\bar{T}<k^n>_{tr},\ppp
<R(k)k^n>_{inc}=(-1)^n\bar{R}<k^n>_{ref}.
\end{equation}
As a consequence, the next correlation is obvious to be valid
\begin{equation} \label{212}
<k^n>_{inc}=\bar{T}<k^n>_{tr}+\bar{R}<(-k)^n>_{ref}.
\end{equation}

\subsection*{The expectation values of the operator $\hat{x}$}

We begin again with the expressions to be common for all three packets.
We have
\begin{equation} \label{213}
<\Psi|\hat{x}|\Psi>=\frac{1}{2\pi} \intx \intk dk'
f^*(k',t)f(k,t)x\exp[i(k-k')x]
\end{equation}
Substituting $-i\frac{\da}{\da k}\exp(i(k-k')x)$ for the expression
$x\exp(i(k-k')x)$, and integrating in parts, we find that
\[<\Psi|\hat{x}|\Psi>=i\intk f^*(k,t)\frac{\da f(k,t)}{\da k}= \]
\begin{equation} \label{214}
= i\intk M\aro\frac{dM\aro}{dk} -\intk M^2\aro\frac{\da\xi(k,t)}{\da k}.
\end{equation}
Since the first term here is equal to
\[\frac{i}{2}M^2\aro|^{+\infty}_{-\infty}=0,\]
we have
\begin{equation} \label{215}
<\Psi|\hat{x}|\Psi>=-\intk M^2\aro \frac{\da\xi(k,t)}{\da k}\equiv
-<f|\frac{\da\xi(k,t)}{\da k}|f>.
\end{equation}

For the incident and transmitted packets, taking into account
expressions (\ref{202}) and (\ref{203}) for $\xi(k,t)$,  we obtain

\begin{equation} \label{216}
<\hat{x}>_{inc}= \frac{\hbar t}{m}<k>_{inc},
\end{equation}
\begin{equation} \label{217}
<\hat{x}>_{tr}= \frac{\hbar t}{m}<k>_{tr}-<J'(k)>_{tr}+d.
\end{equation}
Since the functions $J'(k)$ and $F'(k)$ are even, from (\ref{204}) it
follows that
\begin{equation} \label{218}
<\hat{x}>_{ref}=2a+<J'(k)-F'(k)>_{ref}-\frac{\hbar t}{m}<-k>_{ref}.
\end{equation}
Let, at the instant $t,$ $S_{tr}$ be the distance between the CM of the
transmitted packet and the nearest boundary of the barrier, i.e.,
$S_{tr}=<\hat{x}>_{tr}-b$. Similarly, let $S_{ref}$ be the distance
between the CM and the corresponding barrier's boundary for the
reflected packet at the same instant:  $S_{ref}=a-<\hat{x}>_{ref}$.
From (\ref{217}) and (\ref{218}) it follows that

\begin{equation} \label{2170}
S_{tr}(t)= \frac{\hbar t}{m}<k>_{tr}-<J'(k)>_{tr}-a,
\end{equation}
\begin{equation}
\label{2180} S_{ref}(t)=\frac{\hbar
t}{m}<-k>_{ref}-<J'(k)-F'(k)>_{ref}-a.
\end{equation}
Let us define now the average distance $S_{tr+ref}(t)$ describing the
both packets jointly:
\begin{equation} \label{1190}
S_{tr+ref}(t)=\bar{T} S_{tr}(t)+\bar{R}S_{ref}(t).
\end{equation}
Considering (\ref{2170}), (\ref{2180}) and (\ref{212}), we get

\begin{equation} \label{2190}
S_{tr+ref}(t)=\frac{\hbar t}{m}<k>_{inc}-\bar{b},
\end{equation}
where $\bar{b}=a+<J'(k)>_{inc}-<R(k)F'(k)>_{inc}$ (note that
$<J'(k)>_{inc}=d$, and $<F'(k)>_{inc}=0$ when $V(x)=0$).

\subsection*{The mean-square deviations in $x$-space}

Let us derive firstly the expression to be common for all packets. We
have
\[<\Psi|\hat{x}^2|\Psi>=\frac{1}{2\pi} \intx \intk dk'
f^*(k',t)f(k,t)x^2\exp[i(k-k')x].\]
Substituting the expression $-\frac{\da^2}{\da k^2}\exp(i(k-k')x)$ for
that $x^2\exp(i(k-k')x)$, and integrating in parts, we find
\[<\Psi|\hat{x}^2|\Psi>=-\intk f^*(k,t)\frac{\da^2f(k,t)}{\da k^2}. \]
Since
\[\frac{\da^2f(k,t)}{\da k^2}=\left[M''-M(\xi')^2+i(2M'\xi'+M\xi'')
\right]e^{i\xi},\]
we have
\begin{equation} \label{219}
<\Psi|\hat{x}^2|\Psi>=\intk M [M(\xi')^2-M'']-i\intk
[(M^2)'\xi'+M^2\xi'']
\end{equation}
(hereinafter the prime denotes the derivative with respect to $k$ if
the functions of two variables are written without the independent
variables). One can easily show that the last integral in (\ref{219})
is equal to zero.  Therefore

\begin{equation} \label{220} <\Psi|\hat{x}^2|\Psi>=\intk M^2\aro
[\xi'(k,t)]^2 +\intk [M'\aro]^2.
\end{equation}

Let, for any operator $\hat{Q}$, $<(\delta \hat{Q})^2>$ be the
mean-square deviation $<\hat{Q}^2>-<\hat{Q}>^2$; $\delta\hat{Q}
=\hat{Q}-<\hat{Q}>$.  Then for the operator $\hat{x}$
we have
\begin{equation} \label{400}
<(\delta \hat{x})^2>=<(\ln'M)^2>+<(\delta\xi')^2>.
\end{equation}

Now we are ready to determine these quantities for each packet.
Using (\ref{400}) and expressions (\ref{202})-(\ref{204}), one can show
that for incident packet
\begin{equation} \label{221}
<(\delta \hat{x})^2>_{inc}=<(\ln'A)^2>_{inc}+ \frac{\hbar^2t^2}{m^2}
<(\delta k)^2>_{inc}
\end{equation}
(the first term here, in accordance with the initial condition, is
equal to $l_0^2$); for the transmitted packet \[<(\delta
\hat{x})^2>_{tr}=<(\ln'M_{tr})^2>_{tr}+<(\delta J')^2>_{tr}-\]

\begin{equation} \label{222}
-2\frac{\hbar t}{m}<(\delta J')(\delta k)>_{tr}+
\frac{\hbar^2t^2}{m^2} <(\delta k)^2>_{tr};
\end{equation}
for the reflected packet
\[<(\delta \hat{x})^2>_{ref}=<(\ln'M_{ref})^2>_{ref}+<(\delta
J'-\delta F')^2>_{ref}+\]
\begin{equation} \label{223}
+2\frac{\hbar t}{m}<(\delta J'-\delta F')(\delta k)>_{ref}+
\frac{\hbar^2t^2}{m^2} <(\delta k)^2>_{ref}.
\end{equation}

Let us determine now the mean-square value of $(\delta \hat{x})^2$
averaged jointly over the transmitted and reflected packets:

\begin{equation} \label{401}
<(\delta \hat{x})^2>_{tr+ref}=\bar{T}<(\delta \hat{x})^2>_{tr}
+\bar{R}<(\delta \hat{x})^2>_{ref}
\end{equation}
(note that $<(\delta \hat{x})^2>_{tr+ref}\ne<(\delta
\hat{x})^2>_{inc}$ because $<\hat{x}>_{tr}\ne <\hat{x}>_{ref}$ in
the general case.) Considering (\ref{221})- (\ref{223}), we can reduce
this expression to the form

\begin{equation} \label{224}
<(\delta \hat{x})^2>_{tr+ref}=l^2-2\frac{\hbar t}{m}\chi+
\frac{\hbar^2t^2}{m^2} <(\delta k)^2>_{tr+ref},
\end{equation}
where
\[l^2=\bar{T}<(\ln'M_{tr})^2>_{tr} + \bar{R}<(\ln'M_{ref})^2>_{ref}\]
\[+\bar{T}<(\delta J')^2>_{tr}+\bar{R}<(\delta J'-\delta
F')^2>_{ref},\]

\begin{equation} \label{225}
\chi=\bar{T}<(\delta J')(\delta k)>_{tr}+ \bar{R}<(\delta J'-\delta
F')(\delta k)>_{ref},
\end{equation}
\begin{equation} \label{226}
<(\delta k)^2>_{tr+ref}=\bar{T}<(\delta k)^2>_{tr} +\bar{R}<(\delta
k)^2>_{ref}.
\end{equation}

The first two terms in the expression for $l^2$ may be rewritten, using
(\ref{203}), (\ref{204}) and the correlation $T'+R'=0$, as
\[l^2=<(\ln'A)^2>_{inc}-\frac{1}{4}<(\ln'T)(\ln'R)>_{inc}+\]
\begin{equation} \label{227}
+\bar{T}<(\delta J')^2>_{tr}+\bar{R}<(\delta J'-\delta F')^2>_{ref};
\end{equation}
here the second term is positive and not singular both at the resonances
and at the point $k=0$. Since $<(\ln'A)^2>_{inc}=l_0^2,$ we have
$l^2>l_0^2.$

\newpage
\section*{Figure captions}
\noindent Figure 1. The transmission coefficient ($\circ$) as well as
the ratios $<k>_{tr}/k_0$ (solid line) and $<-k>_{ref}/k_0$ (dashed
line) versus $\lg(d/l_0)$, for the rectangular barriers with $V_0=0.3$
eV, $d=5$ nm; $E_0=0.02$ eV, $m=0.067 m_e$; where $m_e$ is the
electron's mass.

\vspace{\baselineskip}

\noindent Figure 2. The spatial delays $<J'-d>_{tr}$ and
$<J'-d>_{ref}$ for the transmitted (solid line) and reflected (dashed
line) particles, respectively.  The barrier's and particle's parameters
are the same as for fig.1.

\vspace{\baselineskip}

\noindent Figure 3. $\ln(\overline{T})$ versus $d$, for $l_0=15$ nm;
the other parameters are the same as for fig.1.

\vspace{\baselineskip}
\noindent Figure 4. The ratios $<k>_{tr}/k_0$ versus $d$, for the same
parameters (see fig.3).

\vspace{\baselineskip}

\noindent Figure 5. The dependence of $<J'>_{tr}$ on $d$, for the same
parameters (see fig.3).

\end{document}